\documentclass{article} % For LaTeX2e
\usepackage{iclr2025_conference,times}

\usepackage{amsmath,amsfonts,bm}

\def\eqref#1{equation~\ref{#1}}
\def\1{\bm{1}}

\DeclareMathAlphabet{\mathsfit}{\encodingdefault}{\sfdefault}{m}{sl}
\SetMathAlphabet{\mathsfit}{bold}{\encodingdefault}{\sfdefault}{bx}{n}

\usepackage{hyperref}
\usepackage{url}

\title{TF-IDF and BM25 Are Exact KL Divergences}

\author{Ivan Silajev \\Independent Researcher}

\iclrfinalcopy % Uncomment for camera-ready version, but NOT for submission.
\begin{document}

\maketitle

\begin{abstract}
TF-IDF and BM25 are two of the most widely used methods for scoring query–document relevance, yet neither has a standard probabilistic derivation that justifies it as a statistical method within a unified framework. We address this gap by showing that both scoring methods admit an exact interpretation as Kullback-Leibler divergences between two probability models. We treat the BM25 variant that includes the $+1$ correction in the IDF term, which is the one used in practice, and also discuss the original BM25 formulation without that correction. The resulting framework provides a common theoretical basis for TF-IDF and BM25, clarifies what they measure, and allows them to be compared theoretically with other information retrieval methods rather than only experimentally.
\end{abstract}

\section{Introduction}

TF-IDF and BM25 are two ways of calculating the relevance of a query to a document from a target corpus \cite{robertson1994okapi} \cite{robertson2009probabilistic} \cite{manning2008introduction}. Both methods use token frequencies and document lengths as key statistics in their calculation. 

Mathematically speaking, suppose that we have a vocabulary $\mathcal{T}$ of tokens and define a document as a function $d : \mathcal{T} \rightarrow \mathbb{N}_0$ where $d\left(t\right)$ for some $t \in \mathcal{T}$ is the total number of times $t$ occurs in document $d$. Then, we define $| d | \mathrel{:=} \sum_{t \in \mathcal{T}} d\left(t\right)$ as the length of the document $d$. Finally, suppose that we have a corpus $C \mathrel{:=} \left(d_i\right)_{i = 1}^N$ of $N \in \mathbb{N}$ documents. We define $\text{DF}\left(t\right) \mathrel{:=} \sum_{i = 1}^N \mathbb{I}\left\{d_i \left(t\right) > 0\right\}$ as the frequency of the token $t \in \mathcal{T}$ in the entire corpus. Formally, the definition of the Term Frequency-Inverse Document Frequency (TF-IDF) scoring method for a given token $t \in \mathcal{T}$ and document index $i \in \left[1 , N\right]_{\mathbb{Z}}$ is as follows:

\[
\text{TF-IDF}\left(t , i\right) \mathrel{:=} \underbrace{\frac{d_i \left(t\right)}{| d_i |}}_{\text{TF}} \cdot \underbrace{\ln\left(\frac{N}{\text{DF}\left(t\right)}\right)}_{\text{IDF}}
\]

Meanwhile, the formal definition of the Okapi Best Matching 25 (BM25) scoring method for a given token $t \in \mathcal{T}$ and document $d$ is as follows, where the term frequency saturation $k_1 \in \left[0 ,\infty\right)$ and document length normalisation $b \in \left[0 , 1\right]$ are modifiable hyper-parameters and $\text{avgdl} \mathrel{:=} \frac{1}{N} \sum_{i = 1}^N | d_i |$ is the average document length:

\[
\text{BM25}\left(t , i\right) \mathrel{:=} \underbrace{\frac{d_i \left(t\right)\left(k_1 + 1\right)}{d_i \left(t\right) + k_1 \left(1 - b + b \cdot \frac{| d_i |}{\text{avgdl}}\right)}}_{\text{TF}} \cdot \underbrace{\ln\left(\frac{N -\text{DF}\left(t\right) + \frac{1}{2}}{\text{DF}\left(t\right) + \frac{1}{2}} + 1\right)}_{\text{IDF}}
\]

The version of BM25 we focus on in this paper is the one used in practice, such as in production-grade document search software \cite{milvus_bm25_docs}. We also discuss the original BM25 formula from \cite{robertson1994okapi} in section \ref{origin-bm25}.

The problem behind these two scoring methods is the lack of an exact mainstream probabilistic derivation of both that would justify them as statistical methods within a unified probabilistic framework rather than as designed heuristics. This is a problem because there is no common theoretical basis on which to compare TF-IDF and BM25, as well as other information retrieval methods similar to them; their practical utility can only be compared empirically.

In this short paper, we demonstrate that each of TF-IDF and BM25 is actually an exact Kullback-Leibler divergence between two specific probability models. The KL divergence has long been used in IR, but to our knowledge, it has not been shown that the exact TF-IDF and BM25 term scores arise as KL divergences under a unified model.

The consequence of the paper is an exact probabilistic model that refines the interpretation of what the two scoring methods measure, since the KL divergence has a precise physical interpretation.

In section \ref{theory-section}, we cover the mathematical theory of KL divergence and how to derive both score functions from it. Finally, in section \ref{interpretation-section}, we give a unified interpretation of what TF-IDF and BM25 represent under our divergence framework, as well as what the original version of BM25 without the $+ 1$ correction in the logarithm stands for.

\section{Theory}
\label{theory-section}

\subsection{Kullback-Leibler Divergence}

Given a measurable space $\left(\Omega , \mathcal{F}\right)$, where $\Omega$ is the state space and $\mathcal{F}$ is the event space, a sigma algebra on subsets of $\Omega$, and two probability measures $\mathbb{Q}$ and $\mathbb{P}$ defined on it such that $\mathbb{P}\left(A\right)= 0 \Rightarrow \mathbb{Q}\left(A\right) = 0$ for all $A \in \mathcal{F}$ (i.e. $\mathbb{Q}$ is absolutely continuous with respect to $\mathbb{P}$), the Kullback-Leibler divergence of $\mathbb{Q}$ with respect to $\mathbb{P}$ is given by the following formula:

\[
D\left[\mathbb{Q}\|\mathbb{P}\right] \mathrel{:=} \mathbb{E}_{\mathbb{Q}}\left[\ln\left(\frac{d \mathbb{Q}}{d \mathbb{P}}\right)\right]
\]

Where $\frac{d \mathbb{Q}}{d \mathbb{P}} : \Omega \rightarrow \left[0 ,\infty\right)$ is the $\mathbb{P}$-almost surely unique Radon-Nikodym derivative of $\mathbb{Q}$ with respect to $\mathbb{P}$. When working with laws of random variables, we use specialised notation. Suppose $X$ and $Y$ are random variables on $\left(\Omega , \mathcal{F}\right)$, then we define $\mathbb{Q}_X \mathrel{:=} \mathbb{Q} \circ X^{- 1}$ for any probability measure $\mathbb{Q}$ and $D_X \left[\mathbb{Q}\|\mathbb{P}\right] \mathrel{:=} D\left[\mathbb{Q}_X\|\mathbb{P}_X\right]$. For conditional divergences, we adopt the following notation: $D_{X | Y}\left[\mathbb{Q}\|\mathbb{P}\right]\left(y\right) \mathrel{:=} D\left[\mathbb{Q}_{X | Y = y}\|\mathbb{P}_{X | Y = y}\right]$. These notational definitions can be intuitively extended for multiple variables. Finally, we mention the chain rule for divergences:

\[
D_{X , Y}\left[\mathbb{Q}\|\mathbb{P}\right] = \mathbb{E}_{\mathbb{Q}}\left[D_{X | Y}\left[\mathbb{Q}\|\mathbb{P}\right]\left(Y\right)\right] + D_Y \left[\mathbb{Q}\|\mathbb{P}\right]
\]

The divergence $D\left[\mathbb{Q}\|\mathbb{P}\right]$ has the physical interpretation of being the average number of bits of information gained when your initial guess about the system $\mathbb{P}$ changes to $\mathbb{Q}$, given that $\mathbb{Q}$ is the actual probability distribution of the system \cite[p.~98]{mackay2003information}.

\subsection{Term Frequency-Inverse Document Frequency}

To define TF-IDF as an exact divergence, we first define four random variables on $\left(\Omega ,\mathcal{F}\right)$. 

\begin{enumerate}
 \item The document index: $I : \Omega \rightarrow \left[1 , N\right]_{\mathbb{Z}}$
 \item The token: $T : \Omega \rightarrow \mathcal{T}$
 \item The event $X \mathrel{:=} \left\{T \text{ is a key token of document } I\right\} \in \mathcal{F}$
 \item The event $Y \mathrel{:=} \left\{T \text{ is a common token}\right\} \in \mathcal{F}$
\end{enumerate}
Next, we state that $\mathbb{Q}_{X , T , I} = \mathbb{P}_{X , T , I}$. We set $X$ and $Y$ independent under $\mathbb{Q}$ and set:

\[
\mathbb{P}\left(Y | X^c , T = t , I = i\right) = \mathbb{Q}\left(Y | T = t , I = i\right) = 1
\]

We finally derive the following conditional divergence:

\[
D_{Y , X | T , I}\left[\mathbb{Q} \|\mathbb{P}\right]\left(t , i\right) = - \mathbb{Q}\left(X | T = t , I = i\right) \cdot \ln(\mathbb{P}\left(Y | X , T = t , I = i\right))
\]

From here, we can confidently set our probabilities: 

\begin{itemize}
 \item $\mathbb{Q}\left(X | T = t , I = i\right) \mathrel{:=} \frac{d_i \left(t\right)}{| d_i |}$
 \item $\mathbb{P}\left(Y | X , T = t , I = i\right) \mathrel{:=} \frac{\text{DF}\left(t\right)}{N}$
\end{itemize}
We derive TF-IDF exactly.

\[
D_{Y , X | T , I}\left[\mathbb{Q} \|\mathbb{P}\right]\left(t , i\right) = \text{TF-IDF}\left(t , i\right)
\]

We can set $\mathbb{Q}\left(T = t | I = i\right)$ to any query distribution, such as $\frac{q\left(t\right)}{| q |}$, where $q : \mathcal{T} \rightarrow \mathbb{N}_0$ is a query document, which gives us the total relevance score of some query to the document $i$:

\[
\text{Score}\left(q , i\right) \mathrel{:=} D_{Y , X , T | I}\left[\mathbb{Q}\|\mathbb{P}\right]\left(i\right) = \sum_{t \in \mathcal{T}} \left(\frac{q\left(t\right)}{| q |} \cdot \text{TF-IDF}\left(t , i\right)\right)
\]

\subsection{Okapi Best Matching 25}

To define BM25 as an exact divergence, we use the exact same four variables as for TF-IDF and enforce the same probabilistic constraints between them. However, this time we set the following:

\[
\mathbb{P}\left(Y | X , T = t , I = i\right) \mathrel{:=} \frac{\text{DF}\left(t\right) + \frac{1}{2}}{N + 1}
\]

The negative log of this probability is exactly the IDF term in BM25:

\[
-\ln(\mathbb{P}\left(Y | X , T = t , I = i\right)) = \ln\left(\frac{N -\text{DF}\left(t\right) + \frac{1}{2}}{\text{DF}\left(t\right) + \frac{1}{2}} + 1\right)
\]

To define $\mathbb{Q}\left(X | T = t , I = i\right)$, we use a surrogate probability model $\mathbb{F}$ on the same measurable space $\left(\Omega , \mathcal{F}\right)$. Let $\mathbb{Q}\left(X | T = t , I = i\right) = \mathbb{F} \left(X | T = t , I = i , Z\right) = \mathbb{F}_t \left(X | I = i , Z\right)$, where the event $Z$ is defined as $Z \mathrel{:=} \left\{\text{Randomly picked token from } I \text{ is } T \text{ when it has } T\right\} \in \mathcal{F}$. Then we use Bayes rule on this conditional law to re-express it as follows:

\[
\mathbb{F}_t \left(X | I = i , Z\right) = \frac{\mathbb{F}_t \left(Z , I = i | X\right) \cdot \mathbb{F}_t \left(X\right)}{\mathbb{F}_t \left(Z , I = i | X\right) \cdot \mathbb{F}_t \left(X\right) + \mathbb{F}_t \left(Z , I = i | X^c\right) \cdot \mathbb{F}_t \left(X^c\right)}
\]

We set the following: 

\begin{itemize}
 \item $\mathbb{F}_t \left(X\right) \mathrel{:=} \frac{1}{k_1 + 1}$
 \item $\mathbb{F}_t \left(I = i | X\right) \mathrel{:=} \frac{1}{N}$
 \item $\mathbb{F}_t \left(Z | I = i , X\right) \mathrel{:=} \frac{d_i \left(t\right)}{| d_i |}$
 \item $\mathbb{F}_t \left(I = i | X^c\right) = \mathbb{F}_t \left(I = i | W , X^c\right) \cdot \mathbb{F}_t \left(W | X^c\right) + \mathbb{F}_t \left(I = i | W^c , X^c\right) \cdot \mathbb{F}_t \left(W^c | X^c\right)$
 \item $\mathbb{F}_t \left(Z | I = i , X^c\right) \mathrel{:=} \frac{\mathbb{I}\left\{d_i \left(t\right) > 0\right\}}{| d_i |}$
 \item $\mathbb{F}_t \left(W | X^c\right) \mathrel{:=} b$
 \item $\mathbb{F}_t \left(I = i | W , X^c\right) \mathrel{:=} \frac{| d_i |}{\sum_{j = 1}^N | d_j |}$
 \item $\mathbb{F}_t \left(I = i | W^c , X^c\right) \mathrel{:=} \frac{1}{N}$
\end{itemize}
Where the event $W$ is defined as $W \mathrel{:=} \left\{\text{Document } I \text{ is verbose}\right\} \in \mathcal{F}$. This substitution yields us the following probability, assuming $d_i \left(t\right) > 0$:

\begin{align}
\mathbb{F}_t \left(X | I = i , Z\right) & = \frac{\frac{d_i \left(t\right)}{| d_i |} \cdot \frac{1}{N} \cdot \frac{1}{k_1 + 1}}{\frac{d_i \left(t\right)}{| d_i |} \cdot \frac{1}{N} \cdot \frac{1}{k_1 + 1} + \frac{1}{| d_i |} \cdot \frac{k_1}{k_1 + 1} \cdot \left(\left(1 - b\right) \frac{1}{N} + b \cdot \frac{| d_i |}{\sum_{j = 1}^N | d_j |}\right)} \\
 & = \frac{d_i \left(t\right)}{d_i \left(t\right) + k_1 \left(1 - b + b \cdot \frac{| d_i |}{\text{avgdl}}\right)}
\end{align}

We also assume that $\mathbb{F}_t \left(X | I = i , Z\right) = 0$ if $d_i \left(t\right) = 0$. Multiplying this probability by $k_1 + 1$ gives exactly the TF term in BM25. We derive BM25 exactly up to a scalar multiple.

\[
D_{Y , X | T , I}\left[\mathbb{Q} |\mathbb{P}\right]\left(t , i\right) = \frac{1}{k_1 + 1} \cdot \text{BM25}\left(t , i\right)
\]

Using the query distribution defined in the TF-IDF section, we get the following total relevance score of some query to the document $i$:

\[
\text{Score}\left(q , i\right) \mathrel{:=} D_{Y , X , T | I}\left[\mathbb{Q}\|\mathbb{P}\right]\left(i\right) = \sum_{t \in \mathcal{T}} \left(\frac{q\left(t\right)}{| q |} \cdot \frac{\text{BM25}\left(t , i\right)}{k_1 + 1}\right)
\]

\section{Interpretation}
\label{interpretation-section}

\subsection{Score function}

The score function $\text{Score}\left(q , i\right)$ for both TF-IDF and BM25 indicates how much the query document $q$ consists of uncommon keywords contained in document $i$. In terms of the divergence, it shows how strongly does the guess $\mathbb{P}$ that a keyword status affects its rarity differ from the naive assumption $\mathbb{Q}$ that keywords are just as common as any word.

When the query consists of rare keywords from the target document, the query is very relevant to the document and the score is high. Otherwise, the score is lower.

\subsection{Events}

We deliberately made the interpretations of the events $X , Y$ and $W$ arbitrary, but as logically defined as possible given their probabilistic values, as the original mathematical frameworks for TF-IDF and BM25 do not give concrete definitions for what it means for a word to be a keyword, to be common and for a document to be verbose.

\subsection{Original BM25 definition}
\label{origin-bm25}

The IDF term of the original definition of BM25 is given by the following expression:

\[
\text{IDF}_{\text{Original}} \mathrel{:=} \ln\left(\frac{N -\text{DF}\left(t\right) + \frac{1}{2}}{\text{DF}\left(t\right) + \frac{1}{2}}\right)
\]

This term is achieved by defining another probability model $\mathbb{P}'$ on $\left(\Omega , \mathcal{F}\right)$. This model is the adversary to $\mathbb{P}$ and differs from the original in that it sets the following counter-intuitive probability:

\[
\mathbb{P}'\left(Y | X , T = t , I = i\right) \mathrel{:=} \frac{N - \text{DF}\left(t\right) + \frac{1}{2}}{N + 1}
\]

Then we redefine our score function as the following adversarial divergence form:

\[
\text{Score}\left(q , i\right) \mathrel{:=} D_{Y , X , T | I}\left[\mathbb{Q}\|\mathbb{P}\right]\left(i\right) - D_{Y , X , T | I}\left[\mathbb{Q}\|\mathbb{P}'\right]\left(i\right) = \mathbb{E}_{\mathbb{Q}}\left[\ln\left(\frac{d \mathbb{P}'_{Y , X , T | I = i}}{d \mathbb{P}_{Y , X , T | I = i}}\right)\right]
\]

This new scoring function aligns with the original BM25 scoring method \cite{robertson1994okapi} and can be interpreted as a measure of relevance that not only challenges the relationship between keyword status and rarity, but also the counter-intuitive hypothesis that keyword status decreases word rarity represented by the adversary probability model $\mathbb{P}'$.

\bibliography{iclr2025_conference}
\bibliographystyle{iclr2025_conference}

\end{document}